\documentclass[12pt]{iopart}
\usepackage{iopams}
\usepackage{graphicx}
\usepackage{lineno}
\begin{document}
\title{Densities, isobaric thermal expansion coefficients and isothermal compressibilities of linear alkylbenzene}
\author{X Zhou$^1$, Q M Zhang$^2$, Q Liu$^3$, Z Y Zhang$^1$, Y Y Ding$^4$, L Zhou$^4$ and J Cao$^4$}
\address{$^1$ Hubei Nuclear Solid Physics Key Laboratory, Key Laboratory of Artificial Micro- and Nano-structures of Ministry of Education, and School of Physics and Technology, Wuhan University, Wuhan 430072, China}
\address{$^2$ Department of Nuclear Science and Technology, School of Energy and Power Engineering, Xi'an Jiaotong University, Xi'an 710049, China}
\address{$^3$ School of Physics, University of Chinese Academy of Sciences, Beijing 100049, China}
\address{$^4$ Institute of High Energy Physics, Chinese Academy of Sciences, Beijing 100049, China}
\ead{xiangzhou@whu.edu.cn}
\begin{abstract}
We report the measurements of the densities of linear alkylbenzene at three temperatures over 4 to 23$^\circ$C with pressures up to 10 MPa. The measurements have been analysed to yield the isobaric thermal expansion coefficients and, so far for the first time, isothermal compressibilities of linear alkylbenzene. Relevance of results for current generation (i.e. Daya Bay) and next generation (i.e. JUNO) large liquid scintillator neutrino detectors are discussed.
\end{abstract}

\maketitle

\section{Introduction}
The linear alkylbenzene (LAB), with a formula of C$_6$H$_5$C$_n$H$_{2n+1}$ (n = 10-13), is the solvent of liquid scintillator (LS) in large liquid scintillator detectors, such as Daya Bay\cite{AnNIM,Ding,AnPRL,AnCPC,An2014}, RENO\cite{Ahn} and JUNO (formerly called Daya Bay II)\cite{Wang,LiPRD,LiPIC}. It has been reported that LAB would be the preferred solvent for a large-volume detector because of its high transparency\cite{Wurm}. The precision measurement for the density of LAB is important to determine the target mass of detector and then the total number of free protons (hydrogen nuclei). The uncertainty of the number of proton comes from the target mass and the hydrogen fraction in the target. Daya Bay (and RENO and Double Chooz) determines $\theta_{13}$ by relative measurements with near and far detectors at different baselines. Therefore, the uncertainty from the hydrogen fraction will cancel out. The uncertainty from the target mass will impact on the $\theta_{13}$ measurement directly. Very careful target mass measurements were conducted in Daya Bay, with a specialized 20-ton filling tank equipped with load cells and Coriolis flow meters, as described in Ref. \cite{AnNIM}. The precision is 0.02\%. The number of proton (and consequently the absolute efficiency) has negligible impact to the $\theta_{13}$ measurement. But it is important in other studies such as measurement of the reactor neutrino flux (and consequently the so-called reactor anomaly phenomenon). It is also necessary to obtain the isobaric thermal expansion coefficient and isothermal compressibility of LAB to accurately describe the relations between the density of LAB with temperature and pressure. JUNO detector will locate at 700 m underground. The rock temperature has measured to be 31$^\circ$C. The temperature of the underground hall and the big water pool that contains the liquid scintillator detector will be controlled at 20$\pm$1$^\circ$ C during operation. The density of LAB changes with temperature which results in the expansion or shrink of the volume for the liquid scintillator of a given mass. The detector will be filled 100\% for uniform response. A overflow tank will accommodate the liquid due to thermal expansion. It is important to estimate the overflow volume of the central detector of JUNO by using the isobaric thermal expansion coefficient. The density of LAB also changes with pressure. The largest pressure difference for LAB would be about 0.3 MPa for the central detector of JUNO whose diameter would be about 35 m. The isothermal compressibility of LAB, which had not been measured yet, is imperative for the calculation of the Rayleigh scattering lengths of liquid scintillators\cite{Wurm,Wahid,Seidel,Yeo,Chan}. 

In this paper, we report the measurements for the densities of LAB from 4 to 23$^\circ$C and at pressures up to 10 MPa. We have derived the isobaric thermal expansion coefficients and the isothermal compressibilities of LAB from the data of densities.

\section{Experimental method}
The compressed liquid density of LAB was measured by the vibrating tube method and the whole measurement system was developed by the Thermodynamic Research Group in Xi'an Jiaotong University\cite{Yin,Qiu}. The schematic diagram of the whole measurement system could be found in Ref. \cite{Yin} and \cite{Qiu}. The measurement system contains a tube densimeter Anton Paar DMA-HPM (Dimensions: 210 mm $\times$ 78 mm $\times$ 86 mm) with a measuring cell whose volume is about 2 ml where a U-shaped Hastelloy C-276 tube could be excited electronically to vibrate at the characteristic frequencies of liquid LAB at various temperatures and pressures. The evaluation unit mPDS 2000V3 connected to the tube densimeter could indicate the vibration period with seven significant digits. The tube densimeter was thermostatted by an external thermostatic bath. The measuring cell was insulated from the environment. The temperature of the vibrating tube cell was measured using a 100 $\Omega$ platinum resistance thermometer which had been calibrated over the experimental temperature range against a 25 $\Omega$ reference thermometer certified by the National Institute of Metrology of China. The uncertainty of temperature measurement was estimated to be within $\pm 0.02^\circ$C. The pressure of measurement system was applied by a piston pump HIP 50-5.75-30 and measured by a high pressure transducer HBM P3MB. A data acquisition unit (Agilent 34970A) was employed for the transformation of the pressure transducer measurement signal. The uncertainty of pressure measurement was estimated to be within $\pm 0.06$ MPa. The density measurement system had been calibrated by water and vacuum over the entire temperature and pressure range with the method proposed by Lampreia and Nieto de Castro\cite{Lampreia}. The uncertainty of density measurement for LAB sample was estimated to be within $\pm 0.4$ kg$\cdot$m$^{-3}$.

A vacuum test was applied to the entire system where the pressure had been lower than 10 Pa to ensure the precision and safety of measurements. The entire system had been firstly purged by acetone, and then blown by dry flowing nitrogen gas for three times, and finally purged by the dry flowing nitrogen gas again after one or two days. The process of purge was repeated three times before the measurements. The sample of LAB was provided by the China Jinling Petrochemical Limited Corporation. The attenuation length of the LAB from the same batch is about 20 m\cite{Gao}. It was purified by three freeze-pump-thaw cycles via liquid nitrogen before measurements. A vacuum was applied to the entire pipe circuit to remove air or nonvolatile gases, and then the LAB sample was loaded into the pipe circuit by corresponding valve operations. When the temperature for the vibrating tube had been stable, the vibration period of the U-tube could be determined from the change of pressure between the initial pressure and the maximum pressure.

\section{Results}
The compressed densities of LAB were measured at temperatures 4, 15 and 23$^\circ$C and at pressures 0.1, 0.3, 0.5, 2, 4, 6, 8 and 10 MPa. A total of 24 points were obtained, as listed in table \ref{tab1}. The density of LAB decreases with temperature and increases with pressure. 
\begin{table}[!htp]
\caption{\label{tab1}Experimental densities $\rho$ for LAB at various temperatures and pressures. The uncertainties of temperatures, pressures and densities are 0.02$^\circ$C, 0.06 MPa and 0.4 kg$\cdot$m$^{-3}$, respectively.}
\begin{indented}
\item[]\begin{tabular}{lllllllll}
\br
$p$ [MPa] & 0.10  & 0.30  & 0.50  & 2.00  & 4.00  & 6.00  & 8.00  & 10.00 \\
\mr
\multicolumn{1}{l|}{$T$ [$^\circ$C]} & \multicolumn{8}{c}{$\rho$ [kg$\cdot$m$^{-3}$] } \\
\multicolumn{1}{l|}{\hphantom{0}4.00} & 869.0 & 869.1 & 869.2 & 870.1 & 871.4 & 872.6 & 873.7 & 874.9 \\
\multicolumn{1}{l|}{15.02}            & 860.7 & 860.9 & 861.0 & 861.9 & 863.2 & 864.4 & 865.7 & 866.9 \\
\multicolumn{1}{l|}{22.98}            & 854.5 & 854.7 & 854.8 & 855.8 & 857.1 & 858.4 & 849.6 & 860.9 \\
\br
\end{tabular}
\end{indented}
\end{table}

\subsection{Isobaric thermal expansion coefficients}
\begin{figure}[!htp]
\centering
\includegraphics[width=8cm]{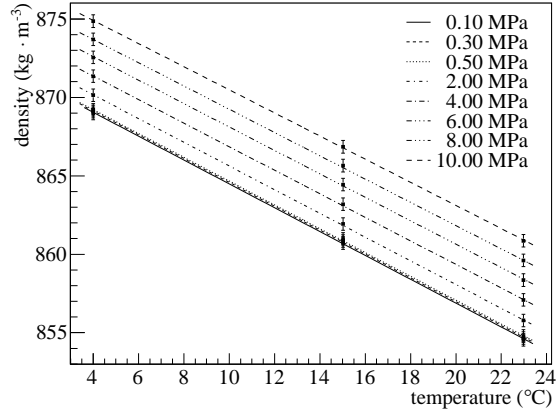}
\caption{\label{fig1}   The fitting curves of $\rho(T)$ at eight pressures.}
\end{figure}

\begin{figure}[!htp]
\centering
\includegraphics[width=8cm]{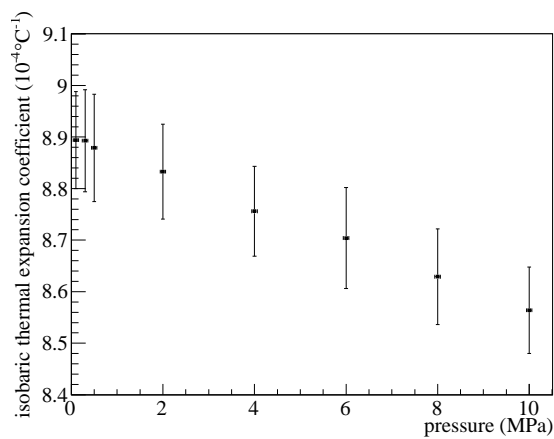}
\caption{\label{fig1-1}   The fitting results of the isobaric thermal expansion coefficients at eight pressures.}
\end{figure}

\begin{figure}[!htp]
\centering
\includegraphics[width=8cm]{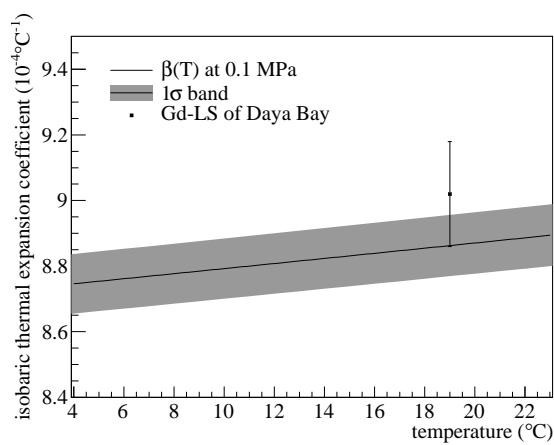}
\caption{\label{fig1-2}   The isobaric thermal expansion coefficients over 4 to 23$^\circ$C at 0.1 MPa with 1$\sigma$ band. Previous measurement of the isobaric thermal expansion coefficient at 19$^\circ$C for Gd-LS of Daya Bay is also shown for comparison.}
\end{figure}

The isobaric thermal expansion coefficient $\beta$ is defined as
\begin{equation}
\beta=\frac{1}{V}\left(\frac{\partial V}{\partial T}\right)_p=-\frac{1}{\rho}\left(\frac{\partial \rho}{\partial T}\right)_p,\label{beta}
\end{equation}
where $V$ is the volume, $p$ is the pressure, $T$ is the temperature and $\rho$ is the density. The isobaric thermal expansion coefficients could be derived from the empirical equation of state
\begin{equation}
\rho(T) = \rho_0[1 - \beta_0 (T - T_0)],\label{eos}
\end{equation}
where $T_0$ is 23$^\circ$C which is the operating temperature for Daya Bay antineutrino detectors, $\beta_0$ is the isobaric thermal expansion coefficient at $T_0$ and $\rho_0$ is the density at $T_0$. The fitting curves are shown in figure \ref{fig1}, the fitting parameters are listed in table \ref{tab2} and $\beta_0$ are shown in figure \ref{fig1-1}.
\begin{table}[!htp]
\caption{\label{tab2}  The fitting parameters for density $\rho(T)$ of LAB.}
\noindent
\scriptsize
\begin{tabular}{llllllllll}
\br
$p$ [MPa]       & 0.10  & 0.30  & 0.50  & 2.00  & 4.00  & 6.00  & 8.00  & 10.00 \\
\mr
$\rho_0$ [kg$\cdot$m$^{-3}$]& 854.6$\pm$0.1   & 854.7$\pm$0.1   & 854.8$\pm$0.1   & 855.8$\pm$0.1   & 857.1$\pm$0.1   & 858.4$\pm$0.1   & 859.6$\pm$0.1   & 860.9$\pm$0.1 \\
$\beta_0$ [10$^{-4}$$^\circ$C$^{-1}$]    & 8.894$\pm$0.094 & 8.893$\pm$0.099 & 8.879$\pm$0.104 & 8.833$\pm$0.092 & 8.756$\pm$0.087 & 8.704$\pm$0.098 & 8.629$\pm$0.093 & 8.564$\pm$0.084 \\
\br
\end{tabular}
\end{table}
The relationship between isobaric thermal expansion coefficient and temperature at a given pressure could be derived by equations (\ref{beta}) and (\ref{eos}) which is
\begin{equation}
\beta(T)=\frac{\beta_0}{1-\beta_0(T-T_0)}.
\end{equation}
By using the parameters in table \ref{tab2}, $\beta(T)$ at 0.1 MPa with 1$\sigma$ band is shown in figure \ref{fig1-2}. 

\subsection{Isothermal compressibility}
\begin{figure}[!htp]
\centering
\includegraphics[width=8cm]{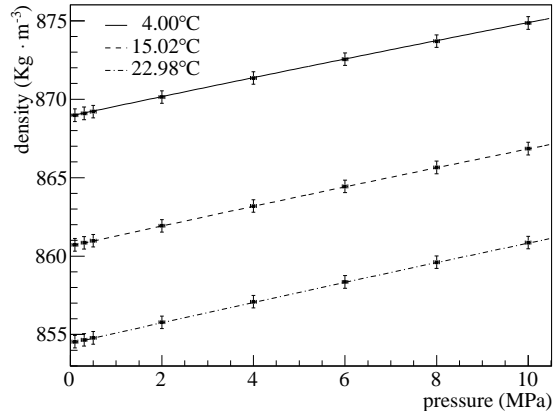}
\caption{\label{fig2}   The fitting curves of $\rho(p)$ at three temperatures. }
\end{figure}

\begin{figure}[!htp]
\centering
\includegraphics[width=8cm]{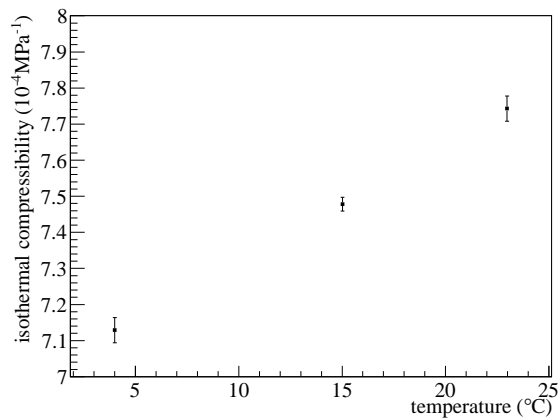}
\caption{\label{fig2-1}   The fitting results of the isothermal compressibilities at three temperatures.}
\end{figure}

\begin{figure}[!htp]
\centering
\includegraphics[width=8cm]{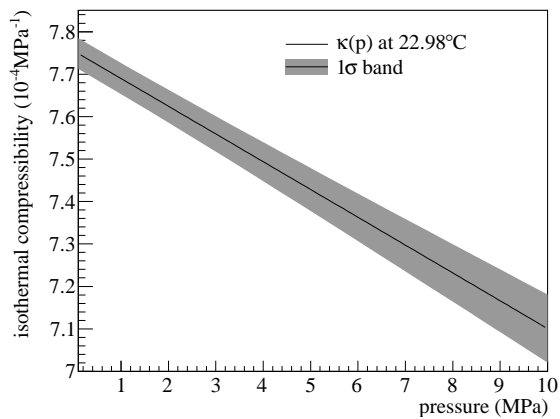}
\caption{\label{fig2-2}   The isothermal compressibilities over 0.1 to 0.5 MPa at 22.98$^\circ$C with 1$\sigma$ band.}
\end{figure}

The isothermal compressibility $\kappa$ is defined as
\begin{equation}
\kappa=-\frac{1}{V}\left(\frac{\partial V}{\partial p}\right)_T=\frac{1}{\rho}\left(\frac{\partial \rho}{\partial p}\right)_T,\label{kappa}
\end{equation}
where $V$ is the volume, $p$ is the pressure, $T$ is the temperature and $\rho$ is the density. The isothermal compressibilities at different temperatures could be derived from the empirical equation of state\cite{Holder}
\begin{equation}
\rho(p) = \frac{\varrho_0}{1 - \kappa_0 (p -p_0) - \kappa_0' (p - p_0)^2},
\label{eos2}
\end{equation}
where $p_0$ is 0.101325 MPa, $\varrho_0$ is the density at $p_0$, $\kappa_0$ is the isothermal compressibility at $p_0$ and $\kappa_0'$ equals to $\frac{1}{2}(\partial\kappa/\partial p)|_{p=p_0}$. The fitting curves are shown in figure \ref{fig2}, the fitting parameters are listed in table \ref{tab3} and $\kappa_0$ are shown in figure \ref{fig2-1}.
\begin{table}[!htp]
\caption{ \label{tab3}  The fitting parameters of $\rho(p)$ for LAB.}
\begin{indented}
\item[]\begin{tabular}{llll}
\br
$T$ [K]                      &4.00             &15.02             &22.98            \\
\mr
$\varrho_0$ [kg$\cdot$m$^{-3}$] &868.974$\pm$0.005  &860.721$\pm$0.003   &854.533$\pm$0.005  \\
$\kappa_0$ [10$^{-4}$MPa$^{-1}$]       &7.129$\pm$0.035  &7.478$\pm$0.019   &7.743$\pm$0.035  \\
$\kappa_0'$ [10$^{-6}$MPa$^{-2}$]  &-3.391$\pm$0.363 &-3.390$\pm$0.194  &-3.258$\pm$0.364 \\
\br
\end{tabular}
\end{indented}
\end{table}
It could be shown in figure \ref{fig2-1} that the isothermal compressibility $\kappa_0$ increases with temperature. The relationship between the isothermal compressibility and pressure at a given temperature could be derived by equations (\ref{kappa}) and (\ref{eos2}) which is
\begin{equation}
\kappa(p)=\frac{\kappa_0+2\kappa_0'(p-p_0)}{1-\kappa_0(p-p_0)-\kappa_0'(p-p_0)^2}.
\label{kappaT}
\end{equation}
By using the parameters in table \ref{tab3}, $\kappa(p)$ at 22.98$^\circ$C with 1$\sigma$ band is shown in figure \ref{fig2-2}. The isothermal compressibility $\kappa(p)$ decrease with pressure since the corresponding $\kappa_0'$ is negative. 

\section{Conclusions}
We measured the densities of LAB at 4, 15, 23$^\circ$C and at 0.1, 0.3, 0.5, 2, 4, 6, 8, 10 MPa. The isobaric thermal expansion coefficients at eight pressures and the isothermal compressibilities at three temperatures were derived by fitting the empirical equations of state. The isobaric thermal expansion coefficient at 19$^\circ$C at 0.1 MPa is consistent with the one for the gadolinium-doped liquid scintillator (Gd-LS) of Daya Bay within the margin of error\cite{AnNIM}. The expansion and contraction of LAB caused by the temperature change per degree around 20$^\circ$C would be about 20 m$^3$ for the 35 m diameter central detector of JUNO. The density, isobaric thermal expansion coefficient and isothermal compressibility of LAB change largely with temperature while they change insignificantly at the pressure difference of 0.3 MPa caused by the large scale of the central detector of JUNO. 

\ack{This work was supported by the Strategic Priority Research Program of the Chinese Academy of Sciences through grant number XDA10010500. We thank the Thermodynamic Research Group (esp. Associate Prof. MENG Xianyang and Dr. FANG Dan) at Xi'an Jiaotong University, which is leaded by Prof. WU Jiangtao, for their help and effort. Z. X. thanks the helpful discussion with Dr. LI Yufeng and Mr. DING Xuefeng. }

\end{document}